# Enhancing the Security of Caesar Cipher Substitution Method using a Randomized Approach for more Secure Communication

Atish Jain
Dept. of Computer Engineering
D.J. Sanghvi College of Engineering
Mumbai University, Mumbai, India

Ronak Dedhia
Dept. of Computer Engineering
D.J. Sanghvi College of Engineering
Mumbai University, Mumbai, India

Abhijit Patil
Dept. of Computer Engineering
D.J. Sanghvi College of Engineering
Mumbai University, Mumbai, India

## ABSTRACT
Caesar cipher is an ancient, elementary method of encrypting plain text message into cipher text protecting it from adversaries. However, with the advent of powerful computers, there is a need for increasing the complexity of such techniques. This paper contributes in the area of classical cryptography by providing a modified and expanded version for Caesar cipher using knowledge of mathematics and computer science. To increase the strength of this classical encryption technique, the proposed modified algorithm uses the concepts of affine ciphers, transposition ciphers and randomized substitution techniques to create a cipher text which is nearly impossible to decode. It also increases the range of characters which Caesar cipher Algorithm can encrypt by including all ASCII and extended ASCII characters in addition to alphabets. A complex key generation technique which generates two keys from a single key is used to provide enhanced security. This paper aims to propose an enhanced version of Caesar cipher substitution technique which can overcome all the limitations faced by classical Caesar Cipher.

## General Terms
Security, Encryption, Ciphers, Cryptography.

## Keywords
Caesar Cipher, Substitution Cipher, Transposition Cipher, Affine Cipher, Encryption, Decryption, Cryptography, Shift Cipher, Randomized, Plain Text, Cipher Text, Cryptanalysis.

## 1. INTRODUCTION
In today's information age, it is impossible to imagine the world without internet. This modern era is dominated by paperless offices mail messages, cash transactions and virtual departmental stores. Large amounts of data is transferred between computers over the internet for professional as well as personal reasons. The computers are interconnected with each other, thus, the communication channels that are used by the computers is exposed to unauthorized access. Hence it has become necessary to secure such data. The obvious defense would be physical security (placing the machine protected behind physical walls). However, due to cost and efficiency issues, physical security is not always a feasible solution. This led to a branch of developing virtual security methods for securing the data from adversaries called as Cryptography. The word cryptography was derived from the Greek word Kryptos, which is used to define anything that is hidden, obscure, secret or mysterious. Cryptography as defined by Yamen Akdeniz is "the science and study of secret writing" that concerns the different ways in which data and communication can be encoded to prevent their contents from being disclosed through various techniques like interception of message or eavesdropping. Different ways include using ciphers, codes, substitution, etc. so that only the authorized people can view and interpret the real message correctly. Cryptography concerns itself with four main objectives, namely, 1) Confidentiality, 2) Integrity, 3) Non-repudiation and 4) Authentication. [1]

Cryptography is divided into two types, Symmetric key and Asymmetric key cryptography. In Symmetric key cryptography a single key is shared between sender and receiver. The sender uses the shared key and encryption algorithm to encrypt the message. The receiver uses the shared key and decryption algorithm to decrypt the message. In Asymmetric key cryptography each user is assigned a pair of keys, a public key and a private key. The public key is announced to all members while the private key is kept secret by the user. The sender uses the public key which was announced by the receiver to encrypt the message. The receiver uses his own private key to decrypt the message.

What is a cipher?

The method of encrypting any text in order to conceal its meaning and readability is called cipher. It is derived from the Arabic word sifr, meaning zero or empty. In addition to ciphers meaning in the context of cryptography, it also means a combination of symbolic letters as in letters for a monogram which is a symbol made by combining two or more letters.[2]

Cryptanalysis refers to the study of ciphers, cipher text, or cryptosystems with a view to finding weaknesses in them that will permit retrieval of the plaintext from the cipher text, without necessarily knowing the key or the algorithm. This is known as breaking the cipher, cipher text, or cryptosystem [3]. The word breaking the cipher can be interchangeably used with the term weakening the cipher. That is to find a property or fault in the design of the encryption algorithm which would help the attacker to reduce the number of keys that he should try while performing brute force attack to break the code. For example, consider a symmetric key algorithm that uses a key of length 2^128 bits which implies that a brute force attack would require the attacker to try all 2^128 possible combinations to be certain of finding the correct key to convert the cipher text into plaintext, which is not possible since it will take thousands of years to try out each and every key. However, a cryptanalysis of the cipher text reveals a method that would allow the plaintext to be found in 2^20 rounds. While it is yet not broken, it is now much weaker and the plaintext can be found with comparatively very less number of tries.

## 2. CLASSICAL METHODS
### 2.1 Caesar Cipher (Substitution technique)

The Caesar cipher is named after Julius Caesar, who, according to Suetonius, used shift cipher with a constant left shift of 3 to encrypt important military messages during the war. Hence it is also known as shift cipher, Caesar's cipher or Caesar shift. It uses a substitution method to evolve the encrypted text. [9] Consider an Example,

Plain text:    ZYXWVUTSRQPONMLKJIHGFEDCBA

Cipher text:   WVUTSRQPONMLKJIHGFEDCBAZYX

When encrypting, an individual looks up each letter of the text message in the "plain text" and writes down the corresponding letter in the "cipher text". Deciphering is done in exactly reverse manner, with a right shift of 3.

This could also be represented using modular arithmetic by transforming the letters into numbers, as per the scheme, a→0, b→1, c→2 … x→23, y→24, z→25. Now, if a letter (x) is to be encrypted, it is expressed as: $En(x) = (x + n) \mod 26$. Decryption is performed similarly: $Dn(x) = (x - n) \mod 26$. The replacement is same for entire text to be encrypted, thus Caesar cipher is classified as monoalphabetic substitution. [4]

The major drawbacks of Caesar cipher is that it can easily be broken, even in cipher-text only scenario. Various methods have been detected which crack the cipher text using frequency analysis and pattern words. One of the approaches is using brute force to match the frequency distribution of letters. This is possible because there are only limited number of possible shifts. (26 in English).

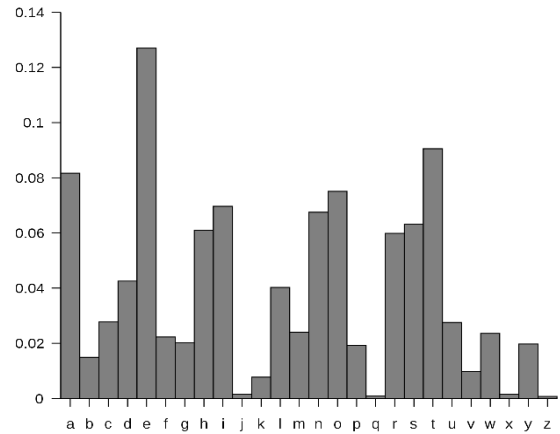

**Figure 1: Typical Frequency Distribution of English Alphabets.**

The distribution of letters in a typical sample of English language text has a very distinct and predictable shape. A Caesar shift "rotates" this distribution, and it is possible to determine the shift by examining the resultant frequency graph. This is the easiest way to break Caesar cipher. Lets take an example to illustrate this weakness. Refer Figure 2 for example.

**Message to encrypt:**
In cryptography, a Caesar cipher, also known as Caesar's cipher, the shift cipher, Caesar's code or Caesar shift, is one of the simplest and most widely known encryption techniques. It is a type of substitution cipher in which each letter in the plaintext is replaced by a letter some fixed number of positions down the alphabet. For example, with a left shift of 3, D would be replaced by A, E would become B, and so on. The method is named after Julius Caesar, who used it in his private correspondence.

**Caesar Output:**                                                              **Shift Value (an integer):** 1260356

Ot ixevzumxgvne, g lgkygx iovnkx, gryu qtuct gy lgkygx'y iovnkx, znk ynolz iovnkx, lgkygx'y iujk ux lgkygx ynolz, oy utk ul znk yosvrkyz gtj suyz cojkre qtuct ktixevzout zkintowaky. Oz oy g zevk ul yahyzozazout iovnkx ot cnoin kgin rkzzkx ot znk vrgotzkdz oy xkvrigikj he g rkzzkx yusk lodkj tashkx ul vuyozouty juct znk grvnghkz. Lux kdgsvrk, cozn g rklz ynolz ul 3, J cuarj hk xkvrgikj he G, K cuarj hkiusk H, gtj yu ut. Znk skznuj oy tgskj glzkx Paroay lgkygx, cnu aykj oz ot noy vxobgzk iuxxkyvutjktik.

**Figure 2: An Example of Encryption using Traditional Caesar Cipher**

This is a text message which is encrypted using a key of 1260356.The encryption technique used is Caesar cipher. The resultant cipher text is also given. Now assume an attacker gets this encrypted cipher text but does not know the key. So to generate a plaintext he tries various cryptanalysis techniques on the cipher text. Suppose, he uses frequency analysis technique to break it. The frequency distribution graph obtained by analyzing this cipher text is shown in Figure 3.

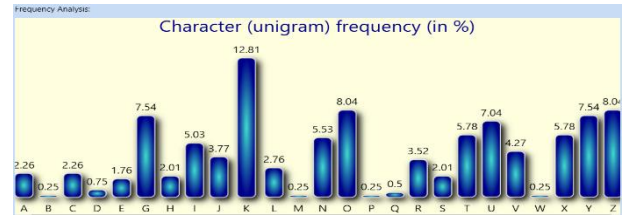

**Figure 3: Frequency Distribution of Characters in Cipher Text.**

**Ciphertext:**
Ot ixevzumxgvne, g lgkygx iovnkx, gryu qtuct gy lgkygx'y iovnkx, znk ynolz iovnkx, lgkygx'y iujk ux lgkygx ynolz, oy utk ul znk yosvrkyz gtj suyz cojkre qtuct ktixevzout zkintowaky. Oz oy g zevk ul yahyzozazout iovnkx ot cnoin kgin rkzzkx ot znk vrgotzkdz oy xkvrigikj he g rkzzkx yusk lodkj tashkx ul vuyozouty juct znk grvnghkz. Lux kdgsvrk, cozn g rklz ynolz ul 3, J cuarj hk xkvrgikj he G, K cuarj hkiusk H, gtj yu ut. Znk skznuj oy tgskj glzkx Paroay lgkygx, cnu aykj oz ot noy vxobgzk iuxxkyvutjktik.

**Deciphered Ciphertext:**                                                          **Key (amount of shift):** 6
In cryptography, a Caesar cipher, also known as Caesar's cipher, the shift cipher, Caesar's code or Caesar shift, is one of the simplest and most widely known encryption techniques. It is a type of substitution cipher in which each letter in the plaintext is replaced by a letter some fixed number of positions down the alphabet. For example, with a left shift of 3, D would be replaced by A, E would become B, and so on. The method is named after Julius Caesar, who used it in his private correspondence.

**Figure 4: Decryption of the Cipher Text using Cryptanalysis Technique of Frequency Analysis.**

On comparing this with the normal frequency distribution of characters in English language (see Figure 1)it is found that the shift is of 6 characters, since 'K' is repeated most in this graph whereas generally 'E' is repeated. Thus, on reverse shifting by 6 characters (as shown in Figure 4), anybody can successfully get the actual plain text and thus the attacker has successfully attacked the system and obtained the actual message.

## 2.2 Columnar Cipher (Transposition cipher)

A transposition cipher rearranges the characters in the plain text to form the cipher text. So, cipher text is obtained by applying permutation on plain text. The key used to decrypt the message is the inverse of the original key.In this method the plain text is written out in fixed length rows, then the columns are rearranged based on the password,then the message is read column by column.If the matrix is not comletely filled then it is padded with null characters or some other character. [5]

Consider an example,

Plain text: enemyattackstonight (we use z to pad empty location).

Key:31452

Step 1: Write the plain text column by column.

```
1  2  3  4  5
e  n  e  m  y
a  t  t  a  c
k  s  t  o  n
i  g  h  t  z
```

Step 2: Rearrange the columns

```
3  1  4  5  2
e  e  m  y  n
t  a  a  c  t
t  k  o  n  s
h  i  t  z  g
```

Step 3:Read the matrix column by column
Cipher text: ettheakimaotycnzntsg

The major drawback of simple Columnar transposition cipher is that to decipher it, the reciever has to work out the column lengths by dividing the length of cipher text by the key length. Then he can write the cipher text out in columns again and then re-order the columns by reforming the key word to obtain the plain text.

## 3. RELATED WORKS

This section describes related research work done to improvise Caesar cipher Encryption Algorithm.

Goyal et al [6] has proposed a modification to the traditional Caesar cipher where he keeps the key size fixed as one. While substitution he checks the index of alphabet, if the index is even then he increases the key value by one else if the index is odd then he decreases the key value by one.

Singh et al. [7] has proposed a technique to combine Caesar cipher with Rail fence transposition technique to eliminate fundamental weaknesses of Caesar cipher and produce a cipher text that is hard to crack.

Omolara et al [8] has proposed a modified hybrid of Caesar cipher and Vigenere cipher to increase the diffusion and confusion characteristics of cipher text by incorporating techniques from modern ciphers like xoring key to the first letter of plain text, xoring first letter of the plain text to second letter and so on.

Disina et al [10] has proposed a method of encryption that depends on the position of the bit in the message. The sender will transpose the bits in the message by shifting the characters in the odd position to the left and characters in the even position to the right side.

Purnama et al [11] has proposed a modified Caesar cipher method where she uses such a method of encryption that the cipher text generated is legible thus making it less suspicious to the cryptanalyst that the text has been encrypted.

## 4. PROPOSED MODIFIED CAESAR ALGORITHM

In this approach, instead of shifting the characters linearly, they are shifted randomly by using the the substitution and permutation box techniques which are implemented in modern encryption techniques like blowfish, DES, etc. Then, substitution box is to be created by implementing the technique of affine ciphers ( i.e. Cipher Text = (Plain Text * key1) + key2). The characters are then replaced by its equivalent values referring to the substitution box. Further, scramble the cipher text to hide the characteristics of the language by using permutation techniques ( i.e. randomly changing the positions of characters in the ciphertext). Permutation of cipher text is done by using double columnar transposition on the cipher text. The proposed algorithm can encrypt the range of characters which the caesar cipher cannot encrypt, i.e. ASCII and Extended ASCII characters.

### 4.1 Algorithm

*4.1.1 Encryption*
1) Create a matrix of N x N.
2) Input the plaintext to be encrypted and a password (key) of size N.
3) From the input password (key) generate two sub keys ckey1 and ckey2 using KeyGenerator() function.
4) Call initialize() function to create two substitution tables.
5) Create ciphertext1 by replacing each character in plaintext by values assigned in the substitution table - map in step 4
6) Perform double columnar transformation using ckey2 which was generated by KeyGenerator() function.
7) Transmit the ciphertext generated in step 6.

*4.1.2 Decryption*
1) Create a matrix of N x N.
2) Input the ciphertext to be decrypted and a password (key) of size N.
3) From the input password (key) generate two sub keys ckey1 and ckey2 using KeyGenerator() function.
4) Call initialize() function to create substitution tables.
5) Perform reverse Columnar transformation twice using ckey2 which was generated by KeyGenerator() function.
6) Generate plaintext by replacing each character in the string obtained by step 5 by values assigned in the substitution table - inversemap in step 4

*4.1.3 KeyGenerator ( ) Function*
1) k1[i]=ASCII value of character in password (key) at position i     // k1 is an array of size N
2) Initialize ckey1=0
3) For i=1 to N do
   a. ckey1 = ckey1*10 + k1[i]
   b. k1[i] = (k1[i] mod 5) + 1
4) Create visited array of size N and initialize it to 0
5) For i = 0 to N do
   a. If k1[i] is not visited mark as visited

b. Else
   i. Find first unvisited value from 1 to N
   ii. Assign it to k1[i] and mark it as visited
6) ckey2=k1

### 4.1.4 Initialize ( ) Function
1) Create two hashmaps - map and inversemap of type character.
2) Assign specific value to a specific key in map as follows
   a. For keys 0 to 32 and 127 to 255 assign value=key
   b. For keys 33 to 126 assign a value using function GetNextRandom ( )
3) Create inversemap by swapping key and their corresponding values from map.

### 4.1.5 GetNextRandom ( ) Function
1) Assign random1 = q mod 127   //Create Global variable random1=0 and q=0
2) q = random1 + 3*ckey1
3) Create mvisited [127] and initialize 0-31 as visited.
4) If random1 is not mvisited mark it as mvisited
5) Else random1 = (random1 + 1) mod 127 and go to step 4.
6) Return the variable random1.

## 4.2 Implementation of Modified Caesar cipher
1) The user enters plaintext and password as input.

   Plaintext: enemy attacks tonight

   Password: qwert

2) Now two subkeys are generated ckey1 and ckey2. Initially the ckey1 is 0.The password 'qwert' is converted into its ASCII equivalent- 113;119;101;114;116.

3) Now the ckey1 is update using the formula.

   ckey1=0*10+113=113

   ckey1=113*10+119=1249

   Continuing like this for all values, the final ckey1 obtained is 1260356.

4) Taking modulus 5 of ASCII equivalent and adding 1 gives ckey2:

   ckey2[1]=4, ckey2[2]=5, ckey2[3]=2

   ckey2[4]=5 but since 5 is already visited, assign it the first not visited value i.e. 1. Therefore, ckey2[4]=1.

   ckey2[5]=2 but again since it is already visited, assign ckey2[4]=3.

   Therefore, ckey1=1260356 and ckey2=45213.

5) Now, generate two substitution tables (hashmap) by calling initialize function(). The two tables are named map and inverse map. Below shown in Table 1 is the map table. Inverse map is just inverse mapping of this table.

**Table 1: Map Table Generated In Above Example.**

| ASCII Value | 32 | 33 | 34 | 35 | 36 | 37 | 38 | 39 | 40 | 41 | 42 | 43 | 44 | 45 | 46 | 47 | 48 | 49 | 50 | 51 | 52 |
|---|---|---|---|---|---|---|---|---|---|---|---|---|---|---|---|---|---|---|---|---|---|
| CHARACTER |   | ! | " | # | $ | % | & | ' | ( | ) | * | + | , | - | . | / | 0 | 1 | 2 | 3 | 4 |
| ENCRYPTED |   | ! | 0 | H | ` | x | " | ) | A | Y | q | # | $ |   | R | j | % | & | 3 | K | c |

| ASCII Value | 53 | 54 | 55 | 56 | 57 | 58 | 59 | 60 | 61 | 62 | 63 | 64 | 65 | 66 | 67 | 68 | 69 | 70 | 71 | 72 | 73 |
|---|---|---|---|---|---|---|---|---|---|---|---|---|---|---|---|---|---|---|---|---|---|
| CHARACTER | 5 | 6 | 7 | 8 | 9 | : | ; | < | = | > | ? | @ | A | B | C | D | E | F | G | H | I |
| ENCRYPTED | { | ' | , | D | \ | t | ( | * | = | U | m | + | - | 6 | N | f | ~ | . | / | G | _ |

| ASCII Value | 74 | 75 | 76 | 77 | 78 | 79 | 80 | 81 | 82 | 83 | 84 | 85 | 86 | 87 | 88 | 89 | 90 | 91 | 92 | 93 | 94 |
|---|---|---|---|---|---|---|---|---|---|---|---|---|---|---|---|---|---|---|---|---|---|
| CHARACTER | J | K | L | M | N | O | P | Q | R | S | T | U | V | W | X | Y | Z | [ | \ | ] | ^ |
| ENCRYPTED | W | 1 | 2 | @ | X | p | 4 | 5 | 9 | Q | i | 7 | 8 | ; | J | b | z | < | > | C | [ |

| ASCII Value | 95 | 96 | 97 | 98 | 99 | 100 | 101 | 102 | 103 | 104 | 105 | 106 | 107 | 108 | 109 | 110 | 111 | 112 |
|---|---|---|---|---|---|---|---|---|---|---|---|---|---|---|---|---|---|---|
| CHARACTER | _ | ` | a | b | c | d | e | f | g | h | i | j | k | l | m | n | o | p |
| ENCRYPTED | S | ? | B | E | T | 1 | F | I | L | M | e | } | O | P | S | ^ | v | V |

6) Create ciphertext1 by replacing each character in plain text by values assigned in map table.
   Plaintext: enemy attacks tonight
   Ciphertext: F^FSk BooBTO] ov^eLMo
7) Now perform columnar on this ciphertext2 using ckey2 we will get

   ```
   1 2 3 4 5
   F ^ F S k
     B o o B
   T O ]   o
   v ^ e L M
   o
   ```

   ```
   4 5 2 1 3
   S k ^ F F
   o B B   o
     o O T ]
   L M ^ v e
             o
   ```

   Ciphertext: So L kBoM ^BO^ F TvoFo]e
8) Again perform columnar on this text

   ```
   1 2 3 4 5
   S o   L
   k B o   M
   ```

| ^ | B | O | ^ |
|---|---|---|---|
| F | T | v | o |
| F | o | ] | e |

| 4 | 5 | 2 | 1 | 3 |
|---|---|---|---|---|
| L |   | o | S |   |
| M |   | B | k | o |
| ^ |   | B | ^ | O |
| v | o |   | F | T |
| e |   | o | F | ] |

Ciphertext: LM^ve  o oBB oSk^FF oOT]

## 4.3 Cryptanalysis of the modified Caesar algorithm

To perform cryptanalysis on this new modified Caesar cipher algorithm take a text message which is encrypted using a key 'qwert'. Modified Caesar cipher algorithm is used for encryption. The resultant cipher text is given as output. The decryption of the cipher text to plain text can be done by using the same key 'qwert'. This example is shown in Figure 5.

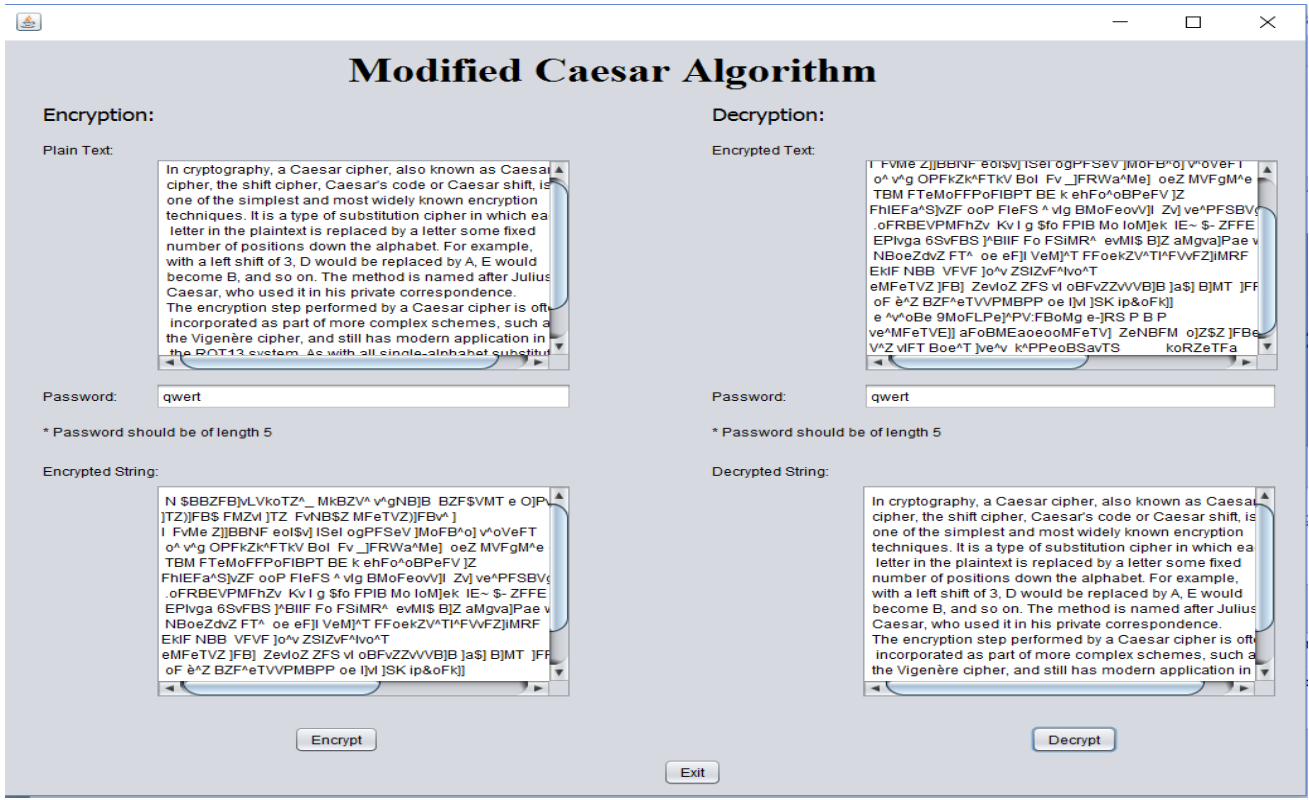

Figure 5: An Example of Encryption using Modified Caesar Cipher

Now, assume an attacker gets this encrypted cipher text but does not know the key. So to generate a plaintext he tries various cryptanalysis techniques on the cipher text. He uses frequency analysis to break the code and decipher the text. The resultant frequency distribution graph obtained by analyzing this cipher text is shown below in Figure 6.

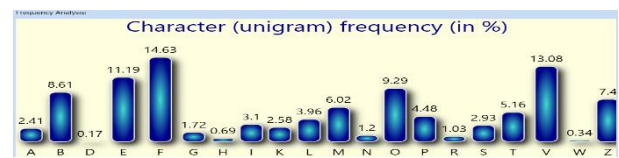

Figure 6: Frequency Distribution of Characters in Cipher Text Generated by Modified Caesar Cipher Approach.

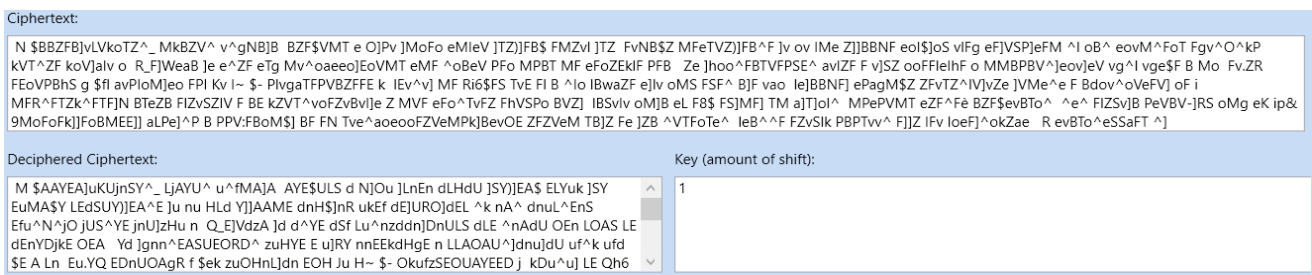

Figure 7: Decryption of the Cipher Text using Cryptanalysis Technique of Frequency Analysis.

On comparing this frequency graph with the normal frequency distribution graph of characters in English language (refer Fig.1)it is found that the shift is of 1 character since 'F' is repeated most in this graph, where generally 'E' is repeated.

Thus, the reverse shifting of 1 character should ideally get the actual plain text to make a successful attack. But looking at the result obtained in Figure 7, it is clear that the obtained plain text is wrong and thus the attacker failed in his attempt to break the cipher text.

Frequency analysis of 100 different samples is done on text encrypted by our modified algorithm and each time it gave an incorrect result. Thus, successfully protecting our text from frequency analysis attack. Furthermore, the characteristics of english language are successfully hidden ( like double and triple words always occuring together, for eg. Is, an, he, she, the, etc.) by using transposition technique, thus, making it even stronger for an attacker to take advantage of language charachteristics to decipher the text. Also, the range of each character in the key is increased to 255, making it impossible to decrypt using brute force attack. Instead of having only 26 no of possible key combinations, the possible combinations of keys is increased to (key length)^256. And it would hence be impossible to use brute force approach to decode, since it will take around millions of years to try out each possible key combination.

## 5. COMPARISION

**Table 2. Comparison of Traditional and Modified Caesar Cipher Algorithm.**

| Traditional Caesar Cipher | Modified Caesar Cipher |
|---|---|
| It linearly shifts all characters by a constant key. | It shifts each character by a random number. (random number is generated by affine cipher) |
| It maintains the characteristics of a language. | It hides the characteristics of a language by spreading the characters throughout the cipher text. |
| It is very easy to implement. | It is comparitively difficult to implement. |
| It is more prone to attacks like frequency analysis attack. | It is not possible to attack using frequency analysis attack. |
| It has many weaknesses which make it easier for a cryptanalyst to attack it. | It overcomes all the weaknesses of traditional caesar cipher making it difficult for a cryptanalyst to attack it. |
| It takes only 26 key combinations to break it using brute force attack. | It takes (key length)^256 possible key combnations to break it using brute force attack. |
| Hence it is very easy to break it using brute force approach. | Hence it is impossible to break it using brute force approach. (since it will take millions of years to try all the possible key combinations) |

## 6. CONCLUSION

In this paper, the limitations and weaknesses of classical encryption algorithms like Caesar cipher and Transposition cipher are described. Then a modified Caesar algorithm is proposed to overcome all the weaknesses and limitations of Caesar cipher. The proposed algorithm uses a randomized approach for substitution which is then combined with double columnar transposition technique to increase the strength. On performing cryptanalysis on the modified algorithm, it is found impossible to break it by frequency analysis. It is practically impossible to decode the algorithm by brute force approach since the attacker would have to try a total of key length raised to 256 different combination of keys. Security provided by this algorithm can be enhanced further by using it with one or more different encryption algorithms or by using asymmetric key approach instead of symmetric key.